\documentclass{aastex631}

\newcommand{\oi}{[O\,{\scriptsize I}]}

\newcommand{\neii}{[Ne\,{\scriptsize II}]}

\newcommand{\uat}[2]{\href{http://vocabs.ands.org.au/repository/api/lda/aas/the-unified-astronomy-thesaurus/current/resource.html?uri=http://astrothesaurus.org/uat/#1}{#2  (#1)}}
\usepackage[T1]{fontenc}

\shorttitle{AASTeX v6.31 Sample article}
\shortauthors{Xie et al.}
\graphicspath{{./}{figures/}}
\usepackage{multirow}
\usepackage{xcolor}

\usepackage[normalem]{ulem}

\begin{document}

\title{Water-Rich Disks around Late M-stars Unveiled: Exploring the Remarkable Case of Sz114}

\newcommand{\affilLPL}{\affiliation{Lunar and Planetary Laboratory, The University of Arizona, Tucson, AZ 85721, USA; \url{cyxie@arizona.edu}}}

\correspondingauthor{Chengyan Xie}
\email{cyxie@arizona.edu}

\author[0000-0001-8184-5547]{Chengyan Xie}
\affilLPL

\author[0000-0001-7962-1683]{Ilaria Pascucci}
\affilLPL

\author[0000-0002-7607-719X]{Feng Long}
\affilLPL
\affiliation{NASA Hubble Fellowship Program Sagan Fellow}

\author[0000-0001-7552-1562]{Klaus M. Pontoppidan}
\affiliation{Jet Propulsion Laboratory, California Institute of Technology, 4800 Oak Grove Dr. Pasadena, CA, 91109, USA}

\author[0000-0003-4335-0900]{Andrea Banzatti}
\affiliation{Department of Physics, Texas State University, 749 N Comanche Street, San Marcos, TX 78666, USA}

\author[0000-0002-5067-1641]{Anusha Kalyaan}
\affiliation{Department of Physics, Texas State University, 749 N Comanche Street, San Marcos, TX 78666, USA}

\author[0000-0003-3682-6632]{Colette Salyk}
\affiliation{Department of Physics and Astronomy, Vassar College, 124 Raymond Avenue, Poughkeepsie, NY 12604, USA}

\author{Yao Liu}
\affiliation{Purple Mountain Observatory, Chinese Academy of Sciences, 2 West Beijing Road, Nanjing 210008, PR China}

\author{Joan R. Najita}
\affiliation{NSF's NOIRLab, 950 N. Cherry Ave., Tucson, AZ 85719, USA}

\author[0000-0001-8764-1780]{Paola Pinilla}
\affiliation{Mullard Space Science Laboratory, University College London, Holmbury St Mary, Dorking, Surrey RH5 6NT, UK}

\author[0000-0003-2631-5265]{Nicole Arulanantham}
\affiliation{Space Telescope Science Institute, 3700 San Martin Drive, Baltimore, MD 21218, USA}

\author[0000-0002-7154-6065]{Gregory J. Herczeg}
\affiliation{Kavli Institute for Astronomy and Astrophysics, Peking University, Yi He Yuan Lu 5, Haidian Qu, 100871 Beijing, China}

\author[0000-0002-6695-3977]{John Carr}
\affiliation{Department of Astronomy, University of Maryland, College Park, MD 20742, USA}

\author[0000-0003-4179-6394]{Edwin A. Bergin}
\affiliation{Department of Astronomy, University of Michigan, Ann Arbor, MI 48109, USA}

\author[0000-0002-4276-3730]{Nicholas P. Ballering}
\affiliation{Department of Astronomy, University of Virginia, Charlottesville, VA 22904, USA}

\author[0000-0002-3291-6887]{Sebastiaan Krijt}
\affiliation{School of Physics and Astronomy, University of Exeter, Stocker Road, Exeter EX4 4QL, UK}

\author[0000-0003-0787-1610]{Geoffrey A. Blake}
\affiliation{Division of Geological and Planetary Sciences, California Institute of Technology, MC 150-21, 1200 E California Boulevard, Pasadena, CA 91125, USA}

\author[0000-0002-0661-7517]{Ke Zhang}
\affiliation{Department of Astronomy, University of Wisconsin–Madison, 475 N. Charter St., Madison, WI 53706, USA}
\affiliation{Department of Astronomy, University of Michigan, 323 West Hall, 1085 S. University Ave., Ann Arbor, MI 48109, USA}

\author[0000-0001-8798-1347]{Karin I. Öberg}
\affiliation{Center for Astrophysics, Harvard \& Smithsonian, 60 Garden St., Cambridge, MA 02138, USA}

\author[0000-0003-1665-5709]{Joel D. Green}
\affiliation{Space Telescope Science Institute, 3700 San Martin Drive, Baltimore, MD 21218, USA}
\affiliation{The University of Texas at Austin, Department of Astronomy, 2515 Speedway, Stop C1400, Austin, TX 78712, USA}

\author{the JDISC collaboration}

\begin{abstract}
We present an analysis of the JDISC JWST/MIRI-MRS spectrum of Sz~114, an accreting M5 star surrounded by a large dust disk with a shallow gap at $\sim 39$\,au. The spectrum  is molecular-rich: we report the detection of water, CO, CO$_2$, HCN, C$_2$H$_2$, and H$_2$. The only identified atomic/ionic transition is  from \neii{} at 12.81\,\micron . A distinct feature of this spectrum is the forest of water lines with the 17.22\,\micron{} emission surpassing that of most mid-to-late M-star disks by an order of magnitude in flux and aligning instead with disks of earlier-type stars. Moreover, flux ratios of C$_2$H$_2$/H$_2$O and HCN/H$_2$O in Sz~114 also resemble those of earlier-type disks, with a slightly elevated CO$_2$/H$_2$O ratio. 
While accretional heating can boost all infrared lines, the unusual properties of Sz~114 could be explained by the young age of the source, its formation under unusual initial conditions (a large massive disk), and the presence of dust substructures. 
The latter delays the inward drift of icy pebbles and help preserve a lower C/O ratio over an extended period. In contrast, mid-to-late M-star disks—which are typically faint, small in size, and likely lack significant substructures—may have more quickly depleted the outer icy reservoir and already evolved out of a water-rich inner disk phase.
Our findings underscore the unexpected diversity within mid-infrared spectra of mid-to-late M-star disks, highlighting the need to expand the observational sample for a comprehensive understanding of their variations and thoroughly test pebble drift and planet formation models.
\end{abstract}



\section{Introduction}
\label{sec:intro}

Observations of disks around young stars provide a unique opportunity to witness planet formation in action \citep[e.g.,][]{Bae2022arXiv221013314B} and  constrain the distribution of volatiles while planets are assembling  \citep[e.g.,][]{Oberg21}. 
{\it Spitzer}/IRS  observations at moderate resolution ($R \sim 700$)  mostly targeted  young solar analogs (FGK and early M) and demonstrated that their disk atmospheres are rich in volatiles. Most of the volatile oxygen and carbon are carried by H$_2$O and CO, while  most nitrogen is likely in the undetected form of gas-phase N$_2$ \citep[e.g.,][and references therein]{Pontoppidan2014prpl.conf..363P}. For the brightest disks, high-resolution ($R \sim 25,000-95,000$) ground-based telescopes targeted few accessible transitions and spectroscopically confirmed that infrared molecular emission mostly traces gas within few au from the star   \citep[e.g.,][]{Pontoppidan10,Mandell2012ApJ...747...92M, Najita2018ApJ...862..122N, Salyk2019ApJ...874...24S, Banzatti23a}.  This is the gas inside the water iceline (hereafter, snowline) that is accreted by those  gaseous planets, including sub-Neptunes, within $\sim 1$\,au \citep[e.g.,][]{Bean2021JGRE..12606639B,Izidoro2022ApJ...939L..19I}. 

In contrast to young solar analogs, only a handful of mid-to-late M-star disks (disks around stars whose spectral types are later than $\sim $M3,)  have IRS spectra and only at modest S/N due to their relatively low luminosity. Nevertheless, their infrared spectra are different and characterized by: i) stronger C$_2$H$_2$ than HCN emission, a signature seen even in lower resolution IRS spectra \citep{Pascucci2009ApJ...696..143P},  and ii) weak/undetected H$_2$O lines \citep{Pascucci2013ApJ...779..178P}.  Retrieval slab models of gas in local thermodynamic equilibrium (LTE) suggest that their inner disk has an enhanced  abundance of C$_2$H$_2$ and HCN relative to water \citep{Pascucci2013ApJ...779..178P}, pointing to elevated C/O ratios of $\sim 1$ \citep[e.g.,][]{Najita2011ApJ...743..147N}. In contrast, when the same models are applied to disks around earlier-type stars, C$_2$H$_2$ and HCN are typically orders of magnitude less abundant than H$_2$O \citep[e.g.,][]{Carr2011ApJ...733..102C,Salyk2011ApJ...731..130S,Walsh15}.  
A similar chemical trend  has been also reported with ALMA for the outer disks ($\ge 20$\,au) of five late M-stars whose C$_2$H/HCN column density ratios are a factor of a few higher than in disks around earlier-type stars\footnote{Unlike C$_2$H$_2$, C$_2$H has a permanent dipole moment hence millimeter transitions.} \citep{Pegues21}. 

These findings hint at a change in the volatile content of mid-to-late M-star disks, with hydrocarbons and nitriles likely being more abundant than in disks around earlier-type stars. 
Supporting this inference, the first JWST/MIRI-MRS spectrum ($R \sim 2,000$) of an early M-star (M1.5, $M_{*} \sim 0.5$\,$M_\odot$, \citealt{Andrews2018ApJ...869L..41A}) disk shows solar-like H$_2$O, C$_2$H$_2$, and HCN abundances \citep[GW Lup,][]{Grant2023ApJ...947L...6G} while that of a late M-star (M5, $M_{*} \sim 0.2$\,$M_\odot$) disk has no water detection but strong emission from C-bearing molecules including from benzene C$_6$H$_6$ \citep[2MASS J16053215-1933159, hereafter J160532,][]{Tabone2023NatAs.tmp...96T}. However, the selection of these two disks was based on the availability of {\it Spitzer}/IRS spectra at $R \sim 700$, which provided the foundation for the aforementioned identified trends. As it is difficult to draw any firm conclusion from such a small sample, targeting disks that have not been previously observed with {\it Spitzer}/IRS is essential, particularly for the underrepresented late M stars, as it allows for a comprehensive exploration of the potential diversity within this class.

Here, we present the JWST/MIRI-MRS spectrum of the disk of Sz~114, an accreting M5 star \citep[$M_* \sim 0.17$ $M_\odot$, e.g.,][]{Hughes1994AJ....108.1071H,Mortier2011MNRAS.418.1194M,Acala17} located in the Lupus~III sub-group \citep[median distance of 158.9\,pc and median age of 2.5\,Myr,][]{Galli2020}. This disk, which is massive \citep[$M_{\rm dust} \sim 30~M_{\oplus}$,][]{Ansdell18}{}{} and millimiter-bright, was observed as part of the ALMA Disk Substructures at High Angular Resolution Project  \citep[DSHARP,][]{Andrews2018ApJ...869L..41A} and found to have a shallow dust gap at $\sim 39$\,au \citep{Huang18}. 
Follow-up visibility modeling using the super-resolution code {\tt frank} suggests additional unresolved substructures between $\approx 7-12$\,au \citep{Jennings22}.
Sz~114 was not observed with {\it Spitzer}/IRS at $R \sim 700$. 
We show here that the mid-IR spectrum of this late M-star disk is remarkably water-rich (Sects.~\ref{sec:obs} and \ref{sec:water-rich}). We discuss molecular flux ratios relative to water in the context of other mid-to-late M-star disks as well as  disks around earlier-type stars observed with {\it Spitzer}/IRS and JWST/MIRI-MRS. We propose that the rich water spectrum of Sz~114 is driven by the star's relatively high accretion luminosity in combination with a young age and a more massive large disk with substructures that might prolong the inward drift of icy pebbles (Sect.~\ref{sec:context}). We conclude in Sect.~\ref{sec:summ} by summarizing the key findings of our study.  Additionally, we emphasize the importance of expanding the sample of mid-to-late M-star disks beyond that observed with {\it Spitzer}/IRS. The extension to systems with a range of ages, accretion luminosities, and outer disk substructures is crucial for capturing the diversity of inner M-star disks and thoroughly testing our current understanding of pebble drift and planet formation.

\section{Observations and data reduction} \label{sec:obs}
Sz~114 was observed with the Medium Resolution spectrometer \citep[MRS,][]{Wells15} of the  JWST Mid-Infrared Instrument \citep[MIRI,][]{Rieke15}  as part of the Cycle~1 Treasury Project ``A DSHARP-MIRI Treasury Survey of Chemistry in Planet-forming Regions''
(ID: 1584, PI: C. Salyk, Co-PI: K. Pontoppidan, hereafter JDISC). To minimize fringes, the source was first acquired using the MRS target acquisition procedure with a neutral density filter. This procedure identifies the brightest pixel in the image and offsets it to the Integral Field Unit's (IFU) center with an accuracy of 0\farcs1. JDISC employed the 4-point dither pattern, which is optimized for point sources, in the negative direction, i.e. the pattern moves from the bottom right to upper left. All three MIRI-MRS grating settings were selected to ensure full spectral coverage ($4.9-28$\,\micron ). To achieve a signal-to-noise (S/N) of $\sim 100$ at the longest wavelengths, 75 groups per ramp were selected for an exposure time per module of 832.5\,s.

\begin{figure*}[htb!]
    \centering
	\includegraphics[width=0.99\textwidth]{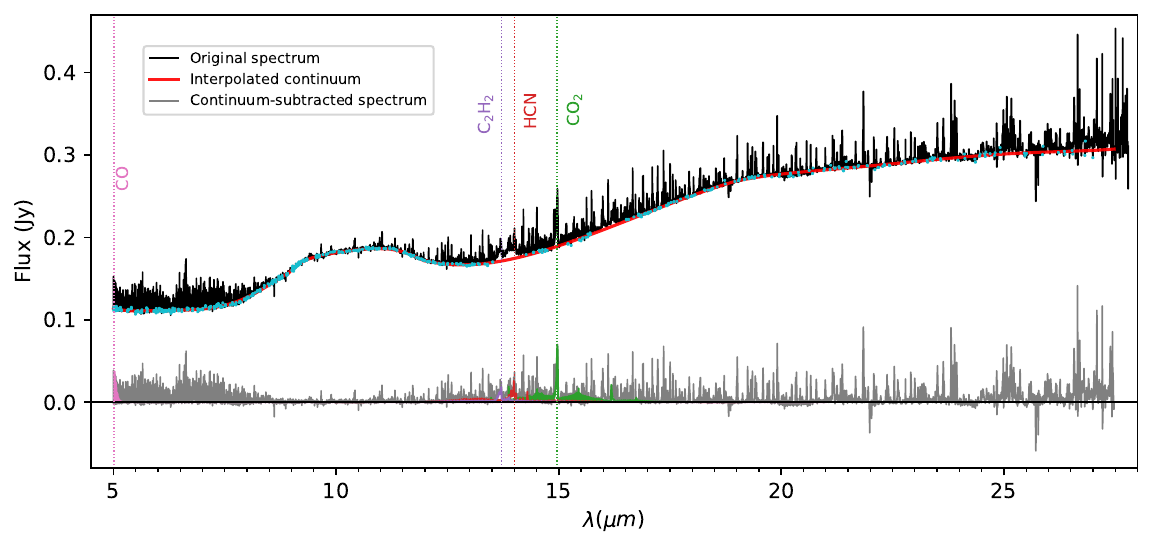}
    \caption{Reduced MIRI-MRS spectrum of Sz~114 (black). The continuum (red line) is interpolated at the light blue dots and superimposed on the original spectrum. The continuum-subtracted spectrum is shown below in grey. The spectrum is dominated by water emission lines, wavelengths of other prominent molecular ro-vibrational bands are indicated as vertical lines. The dips at $\sim 19$, $\sim 22$, and $\sim 26$\,\micron\ are not real but rather data reduction artifacts.
    }
    \label{fig:consub}
\end{figure*}

The data were reduced following  Pontoppidan et al. in prep. In short, the 
JWST Calibration Pipeline \citep{Bushouse22} version 1.11.0 and Calibration Reference Data System (CRDS) context jwst\_1105.pmap  was used to process the data up to stage 2 resulting in datacubes for every exposure, channel, and sub-band. 1D-spectra were extracted with an aperture that widens linearly with wavelength within each sub-band and the background was subtracted from the two opposing dither positions. Then spectra were individually divided by a relative spectral response function constructed from the asteroid  515 Athalia (PID: 1549, PI: K. Pontoppidan) before median-clipping of the four dither positions. Finally, we shifted the spectrum by -1\,km/s to align the lines with the mean heliocentric radial velocity of the Lupus~III subgroup, where Sz~114 is located \citep{Galli13}.
The final reduced spectrum is shown in black in Figure~\ref{fig:consub}.

\begin{deluxetable}{lcc}
\tablecaption{Properties of Sz~114 relevant to this work\label{tab:sourceprop}}
\tablewidth{0pt}
\tablehead{
\colhead{Property (unit)} & \colhead{Value} & \colhead{Ref.} 
}
\startdata
distance (pc) & $156.8\pm0.6$ & 1,3\\
spectral type & M5 & 2 \\
$L_*$ ($L_\odot$) &$0.196$ & 2,3\\
$M_*$ ($M_\odot$) & $0.16$& 2,3\\
$log(\dot{M})$ ($M_{\odot}/yr$) &$-9.1$ & 2,3\\
$log(L_{acc})$ ($L_{\odot}$) &$-2.7$ & 2,3\\
$F_{\rm 0.89\,mm}$ (mJy) & $96.41$ & 4\\
disk inclination ($^\circ$) & 21.3$\pm$1.3 &5 \\
$R_{\rm dust}$ (au) &$56\tablenotemark{a}\pm1$ &5 \\
\enddata
\tablenotetext{a}{Scaled to the Gaia distance in this Table}
\tablerefs{1.~\cite{Gaia3}; 2.~\cite{Acala17}; 3.~\cite{Manara22} ; 4.~\cite{Ansdell16}; 5.~\cite{Huang18}
} 
\end{deluxetable}

\section{The water-rich mid-infrared spectrum of Sz~114}\label{sec:water-rich}
The MIRI-MRS spectrum of Sz~114 shows a broad and flat-topped 10\,\micron{} silicate emission feature typical of mid-to-late M-star disks \cite[e.g.,][]{Apai2005Sci...310..834A,Pascucci2009ApJ...696..143P}. 
These properties indicate that the typical dust grains in the disk atmosphere are  microns in size, i.e. about ten times larger than interstellar medium dust and typical grains in solar-type disk atmospheres \citep[e.g.,][]{Natta07}{}{}.
In addition, the prominent ro-vibrational bands from C$_2$H$_2$, HCN, and CO$_2$, all of which were previously detected with {\it Spitzer}/IRS in mid-to-late M-star disks \citep{Pascucci2009ApJ...696..143P,Pascucci2013ApJ...779..178P}, are also present in the Sz~114 MIRI-MRS spectrum, see Figure~\ref{fig:consub}. At the shortest wavelengths, several CO $\nu=1-0$ P-branch lines are also found, while CO emission in disks has often been found to have non local thermodynamic equilibrium (non-LTE) excitation \citep[e.g.,][]{Thi13,Vanderplas15}{}{}.

The most prominent characteristic of this spectrum is the forest of water lines: from the hot bending ro-vibrational band at $\sim 6.6$\,\micron\  to the cooler pure rotational lines at longer wavelengths (see e.g. Fig.~1 in \citealt{Banzatti23a}). In relation to water emission, Sz~114 is very different from other mid-to-late M-star disks observed with {\it Spitzer}/IRS \citep{Pascucci2013ApJ...779..178P} and the similarly late M5 object recently observed with MIRI-MRS \citep[J160532,][]{Tabone2023NatAs.tmp...96T} in which water emission was either very weak or absent. Such a rich water spectrum is more akin to that of disks around earlier-type stars (spectral types earlier than M3, e.g. \citealt{Carr2011ApJ...733..102C,Salyk2011ApJ...731..130S}), as discussed further in Sect.~\ref{sec:context}. Visual inspection of the MIRI datacubes does not show any extended emission. Therefore, as in previous papers \citep[e.g.,][]{Salyk2011ApJ...731..130S,Pascucci2013ApJ...779..178P}, we assume that the water emission is tracing the disk surface instead of jets or winds. 

Properly identifying weak emission lines from other molecules requires subtracting the dust continuum emission and the rich water spectrum. To identify line-free channels for continuum subtraction, we first generated a water spectrum with CLIcK \citep{Liu2019A&A...623A.106L}. CLIcK assumes a plane-parallel slab of gas in LTE and adopts molecular parameters from the HITRAN database\citep[e.g.,][]{HITRAN22}. Line overlap is properly accounted for using eq.~1 in  \cite{Tabone2023NatAs.tmp...96T}. 
With the generated water spectrum, we identified by eye continuum regions that avoided water lines and the prominent ro-vibrational bands mentioned above (see light blue points in Figure~\ref{fig:consub}). After experimenting fits with polynomials of different orders, we found that a 5th-order polynomial provides a good fit for the entire continuum emission\footnote{We used \texttt{curvefit} within \texttt{scipy} \citep{Pauli20} for the optimization}.
The continuum-subtracted spectrum of Sz~114 is depicted in grey  below the original spectrum in Figure~\ref{fig:consub}.

\begin{figure*}[htb!]
    \centering
	\includegraphics[width=0.99\textwidth]{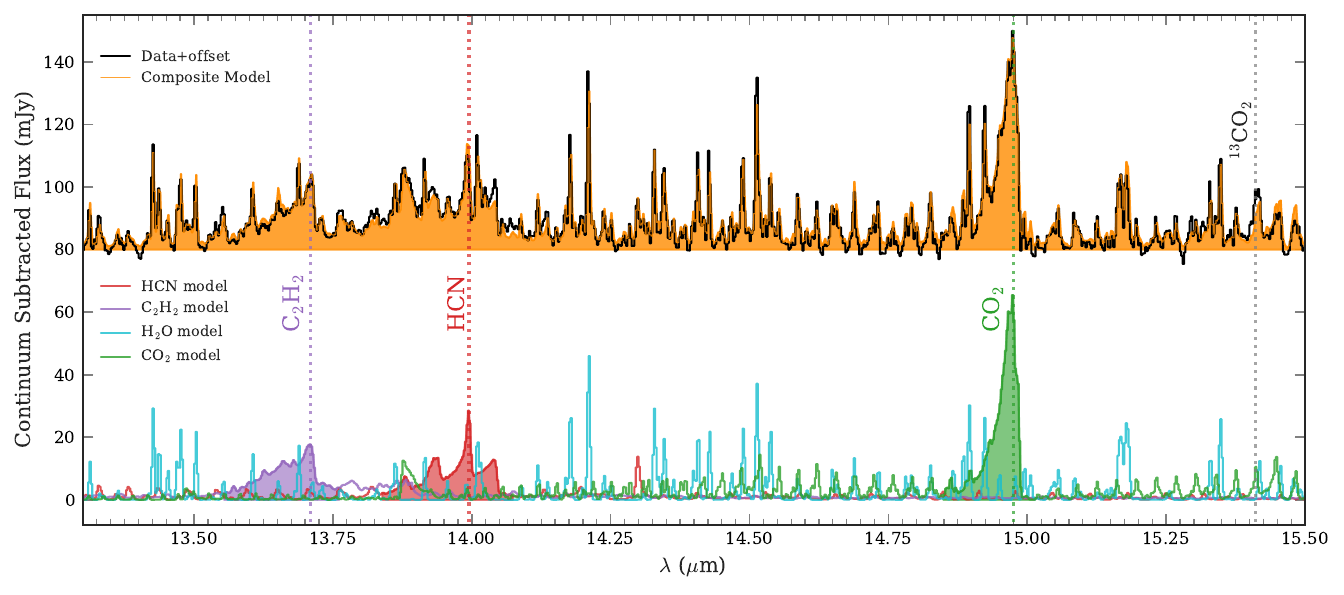}
    \caption{Upper panel: composite model (orange shaded region) between 13.4-15.5\,$\mu$m compared to the Sz~114 JWST-MIRI spectrum (black). Lower panel: individual best fits to  water, C$_2$H$_2$, HCN, and CO$_2$.  
    }
    \label{fig:compositemodel}
\end{figure*}

\begin{figure*}[htb!]
    \centering
	\includegraphics[width=0.99\textwidth]{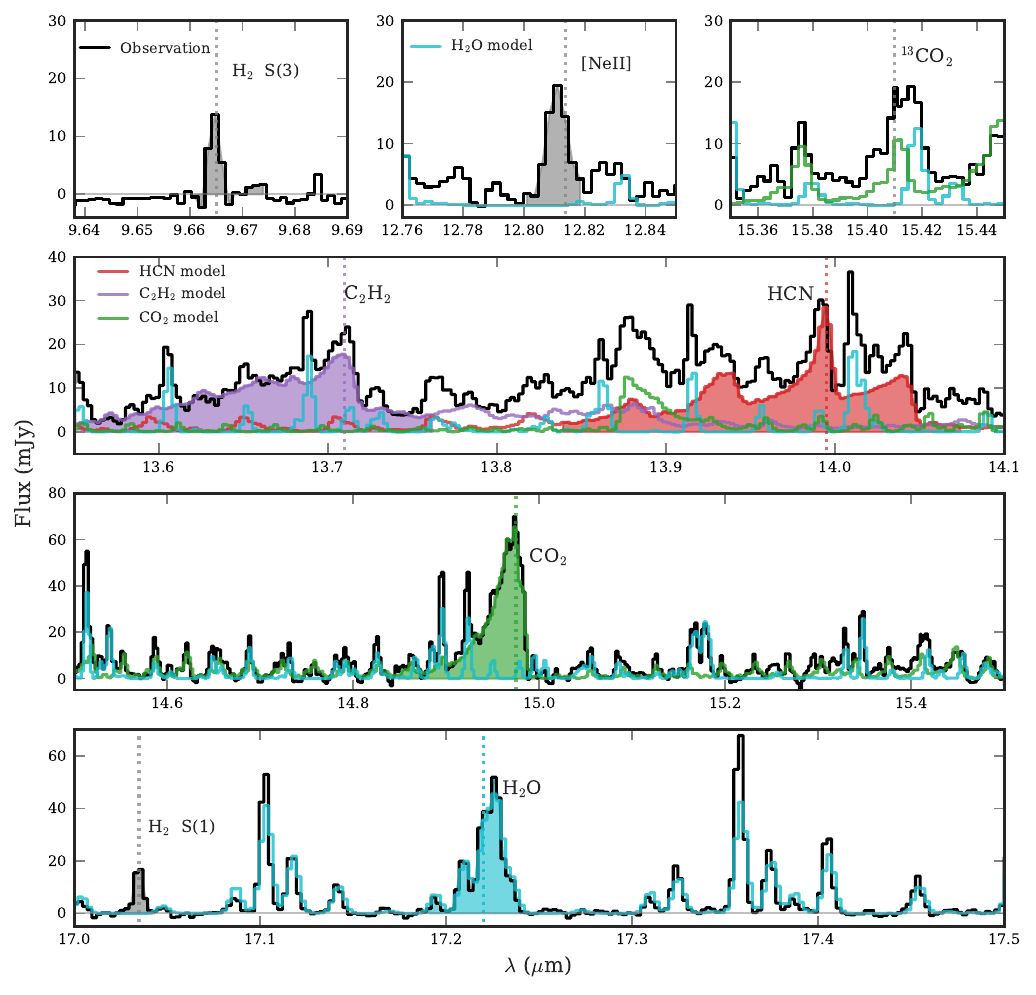}
    \caption{Expanded view of wavelength regions covering the ionic, atomic, and molecular lines reported in Table~\ref{tab:fluxes}. 
    Shaded regions indicate the wavelength ranges over which fluxes are measured. For C$_2$H$_2$ (purple), HCN (red), and CO$_2$ (green)  these are the same ranges as those adopted in earlier {\it Spitzer}/IRS studies to facilitate comparison with literature sources. For H$_2$O (light blue) the wavelength range is restricted to the complex around 17.22\,\micron . The best-fit spectra obtained with CLIcK are also shown with solid lines.
    }
    \label{fig:expandedview}
\end{figure*}

Next, we work with the continuum-subtracted spectrum and run CLIcK again to find  the water models that fit best the  water emission between $\sim 10-19$\,\micron . The reason for focusing on this restricted wavelength range is that it contains the ro-vibrational bands of interest for placing Sz~114 in context of other disks observed with {\it Spitzer}/IRS and, more recently, JWST/MIRI-MRS (see Sect.~\ref{sec:context}).
In spite of the restricted wavelength range, the water spectrum covers a broad range of upper energy levels and Einstein-A coefficients which may not be fitted well by a single-temperature slab of gas in LTE \citep[e.g.,][]{Banzatti2023arXiv230703846B}. This is indeed the case for Sz~114, where we find that two temperatures are necessary to reproduce the observed spectrum (see the sum of the two slab models as light blue lines in the bottom panel of Figure~\ref{fig:compositemodel} and in Figure~\ref{fig:expandedview}). The best fit is found by first fitting the hot water in the region between $\sim 10 - 12$\,\micron{} plus the high energy line at 15.18\,\micron\ ($E_{\rm up} \sim 8,000$\,K) simultanously. Next, the best fit to the hot water is removed and then all water lines are fitted between $\sim 10-12$\,\micron\ and $\sim 16.3-19$\,\micron{} but excluding regions with known molecular lines other than water.  We stress that the aim of this modeling is not to perfectly match the complete JWST-MIRI spectrum. Instead, our goal is merely to identify a suitably precise model within the molecular feature-rich range ($\sim 13-17$\,\micron ) to untangle the contribution from individual molecules and measure their line fluxes.

These slab models have three input parameters: the emitting radius ($R_{\rm em}$, the model fits for an equivalent area of $\pi R_{\rm em}^2$), the column density ($N$), and the gas temperature ($T$).
To fit the water lines, we generate a grid of 4,000 models using a custom-written random walk python code (see appendix~\ref{sec:Appendix:errorbars} for more details) with starting values $R_{\rm em}$=0.2\,au, $N=10^{18}$\,cm$^{-2}$, and $T= 1,000$\,K for the hot and 300\,K for the cool components, based on prior models of solar-type disks \citep{Banzatti23a}. For each step, random values are chosen between $\pm$0.05\,au for $\Delta R_{\rm em,i}$, $\pm$2 for $\ln (\Delta N_i)$, and $\pm$100\,K for $\Delta T_i$. Using equation~2 in \cite{Mulders15} with $M_*$ and $\dot{M}_{\rm acc}$ for Sz~114 (Table~\ref{tab:sourceprop})  and assuming  typical values for the Rosseland mean opacity (from 20 to 570 cm$^2$ g$^{-1}$), a gas-to-dust ratio of 100, and two values for the turbulent mixing strength (0.01 and 0.001), we estimate the midplane snowline to be between $\sim 0.1-0.5$\,au. Since the cool component is still at several hundred K, we set an upper limit of 0.5\,au for $R_{\rm em}$, i.e. most of the water emission is coming within the snowline.  The best-fit model is the one with the minimum $\chi^2$ and derived temperatures, column densities, and emitting radii are summarized in Table~\ref{tab:model} of Appendix~\ref{sec:Appendix:errorbars}. 

After subtracting the water spectrum, we proceed to fit with CLIcK the next major molecular emitters in the $10-19$\,\micron\ region. First, we fit the CO$_2$ spectrum between $14.6-15.4$\,\micron\ thus covering many P- and R-branches in addition to the strongest Q-branch emission, see Figure~\ref{fig:expandedview}. Next, we remove the best-fit spectrum extended to the entire wavelength region and fit together the C$_2$H$_2$ and HCN ro-vibrational bands between $13.5-14.3$\,\micron . The best-fit parameters for these molecular species are also summarized in Table~\ref{tab:model} of Appendix~\ref{sec:Appendix:errorbars} while the composite model compared to the data is shown in the upper panel of Figure~\ref{fig:compositemodel} with individual molecular emission models in the lower panel. 

Aware of the degeneracies among parameters even at the sensitivity and spectral resolution of MIRI-MRS \citep[e.g.][]{Grant2023ApJ...947L...6G}{}{}, we use these fits merely to separate the contribution from different molecules and calculate line fluxes. The fluxes are calculated by integrating the data (or model for the blended molecules C$_2$H$_2$, HCN and CO$_2$) within the wavelength range reported in Table~\ref{tab:fluxes}. To facilitate the comparison between Sz~114 and literature sources these ranges are chosen to match those adopted in {\it Spitzer} studies \citep[e.g.,][]{Salyk2011ApJ...731..130S}.
Regarding water, we note that the two tentative detections in late M-star disks are at 17.22\,\micron\ \citep[in 2MASS~J04381486+2611399 and 2MASS~J04442713+2512164, hereafter J043814 and J044427, see][]{Pascucci2013ApJ...779..178P}, hence we focus on the flux of this line for Sz~114. The calculated line fluxes can be found in Table~\ref{tab:fluxes}. Because the disk surface snowline is further out than the midplane snowline, we have also run models that do not restrict $R_{\rm em}$ to be within $\sim 0.5$\,au. Although the best-fit values for $T$ and $N$ are different from those reported in Table~\ref{tab:model}, molecular line fluxes, which are the focus of this paper, are within 5\% of those reported in Table~\ref{tab:fluxes}. 

Inspection of the continuum-subtracted spectrum with the best-fit water, CO$_2$, C$_2$H$_2$, and HCN synthetic models reveals additional molecular transitions (see Table~\ref{tab:fluxes}): two from H$_2$
and a tentative detection of the ro-vibrational band from the rarer isotopologue $^{13}$CO$_2$ at $\sim 15.41$\,\micron\ (see \citealt{Grant2023ApJ...947L...6G} for a stronger detection in GW~Lup). We have also searched for C$_4$H$_2$, C$_6$H$_6$, and CH$_4$, since they were detected toward J160532 with MIRI-MRS \citep{Tabone2023NatAs.tmp...96T}, but did not identify them in the spectrum of Sz~114.
The only detected atomic/ionic line in the spectrum of Sz~114 is that from \neii\ at 12.81\,\micron\ (see Table~\ref{tab:fluxes}). The H$_2$ peak centroids are consistent with the stellar radial velocity while the \neii\ is clearly blueshifted, see Figure~\ref{fig:expandedview}. A discussion of the H$_2$ and \neii\ lines is provided in Appendix~\ref{app:OtherLines}. Here, we focus on the prominent water lines and the CO$_2$, C$_2$H$_2$, and HCN ro-vibrational bands.  

\begin{deluxetable}{lccc}
\tablecaption{Summary of line fluxes\label{tab:fluxes}}
\tablewidth{0pt}
\tablehead{
\colhead{Species} & \colhead{$\lambda$-range} & \colhead{Flux} &\colhead{Comments}   \\
\colhead{} & \colhead{(\micron)} & \colhead{(10$^{-14}$ erg\,s$^{-1}$\,cm$^{-2}$)}  & \colhead{} 
}
\startdata
\multicolumn{4}{c}{Atomic and Ionic}\\
\hline
\neii\ &12.80-12.82 & 0.3 & \\
\hline
\multicolumn{4}{c}{Molecular}\\
\hline
CO (P35 1-0) & 5.024-5.031 & 1.07 &\\
H$_2$ (0,0) S(3) & 9.66-9.675 & 0.12 &\\
H$_2$ (0,0) S(1) & 17.02-17.045 & 0.07 &\\
C$_2$H$_2$ &13.553-13.764 & 2.12 &a\\
HCN  &13.837-14.075 & 2.23 &a\\
CO$_2$   &14.847-15.014 & 3.61 &a\\
$^{13}$CO$_2$   & 15.39-15.43& 0.19 & b\\
H$_2$O &17.19-17.25 &0.87 &\\
\enddata
\tablecomments{Sz~114 has a forest of water emission lines. We first fit the water spectrum and subtract the best-fit water model before computing the flux from other species or fit the emission from other molecular lines.}
\tablenotetext{a}{Emission from these species is blended, hence  we use our best-fit model flux here.}
\tablenotetext{b}{This is a tentative detection after subtracting blended water and $^{12}$CO$_2$ lines.}
\end{deluxetable}


\section{Sz~114 in the context of other disks}\label{sec:context} 
\begin{figure*}[htb!]
    \centering
	\includegraphics[width=0.99\textwidth]{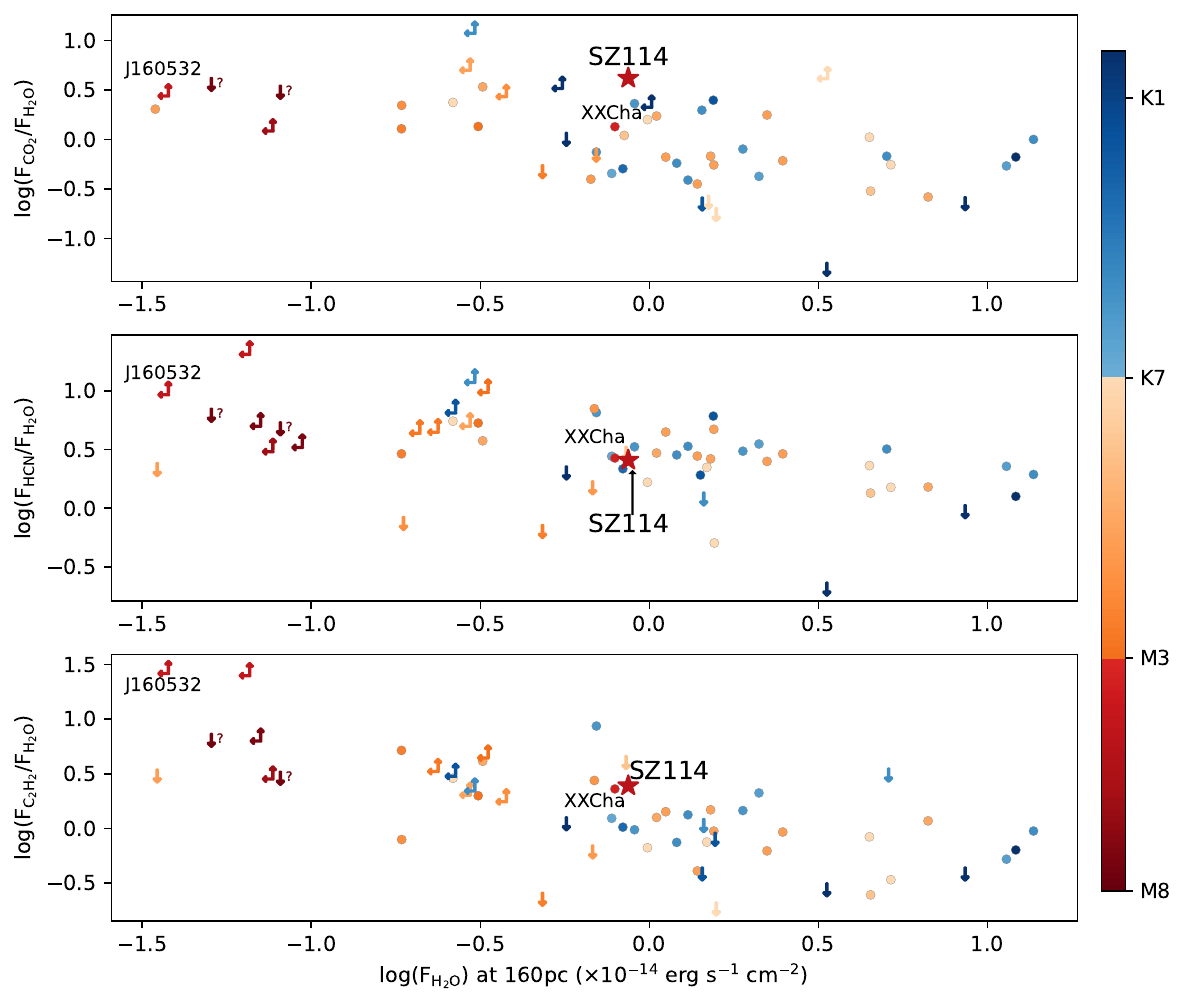}
    \caption{Molecular flux ratios vs. scaled water flux for mid-to-late M-star disks \citep{Pascucci2013ApJ...779..178P} and disks around earlier-type stars \citep{Banzatti2020ApJ...903..124B}. In these comparisons, the water line at 17.22\,\micron\ is adopted. The two downward arrows with a question mark next to them are for the tentative {\it Spitzer} water detections in the mid-to-late M-star disks of J043814 and J044427.  J160532 is a source similar to Sz114 in spectral type, stellar mass and accretion luminosity but is extremely carbon-rich. XX~Cha is a slightly earlier-type (M3.5) star with a disk rich in water. Note how Sz~114 (star symbol) has more similar molecular flux ratios to earlier- than later-SpTy except for CO$_2$/H$_2$O which is higher than the average earlier-type disk value. 
    }
    \label{fig:fluxratio}
\end{figure*}

As described in Sect.~\ref{sec:water-rich}, the MIRI-MRS spectrum of Sz~114 is dominated  by molecular emission, especially from water vapour. As such, its spectrum is very different from that of late M-star disks observed  with {\it Spitzer}/IRS and more similar to that of disks around earlier-type stars. To better illustrate the difference, we show in Figure~\ref{fig:fluxratio} line flux ratios with respect to water emission at 17.22\,\micron , scaled at 160\,pc. We focus on C$_2$H$_2$, HCN, and CO$_2$ as these ro-vibrational bands were previously detected in {\it Spitzer}/IRS and their fluxes or upper limits are tabulated. The sample of late M-type stars is from \cite{Pascucci2013ApJ...779..178P} and includes at most six sources with SpTy  M7.25$-$M4.5 for which line flux ratios (including upper or lower limits) can be calculated. The sample of earlier-type stars (M3$-$G0) is from \cite{Banzatti2020ApJ...903..124B} and includes at most 49 sources\footnote{The source number is different among panels because of non-detections.}. For this latter sample, we note that the water line flux has been re-calculated using the restricted wavelength range around the 17.22\,\micron\ transition. 

Figure~\ref{fig:fluxratio} demonstrates that Sz~114 has an unusually strong 17.22\,\micron{} water flux for its late SpTy compared to other mid-to-late M-star disks observed before, more similar to that of disks around early M and K-type stars. The only other outlier in this sample is the disk of XX~Cha, a slightly earlier SpTy source (M3.5) observed with Spitzer/IRS and more recently with JWST/MIRI but whose MIRI spectrum has not been published yet. Additionally, Sz~114 shares similar HCN/H$_2$O and C$_2$H$_2$/H$_2$O flux ratios with XX~Cha, both of which fall well within the range of values observed in disks around earlier-type stars. One difference is the slightly higher CO$_2$/H$_2$O flux ratio exhibited by Sz~114. In contrast to Sz~114, late M-star disks tend to have higher C$_2$H$_2$/H$_2$O  flux ratios (Figure~\ref{fig:fluxratio} lower panel) and higher C$_2$H$_2$/HCN ratios (see Fig.~13 from \citealt{Pascucci2009ApJ...696..143P}). These higher ratios have been attributed to a very high C/O ratio ($\ge 0.8$) in the inner disk of late M-type stars \citep{Pascucci2013ApJ...779..178P,Tabone2023NatAs.tmp...96T}, as the extra carbon that does not end up in CO can efficiently form hydrocarbons in the gas phase \citep[e.g.,][]{Najita2011ApJ...743..147N}. Sz~114 clearly shows that there is  a  larger diversity of mid-IR spectra down to late spectral types than previously inferred from the {\it Spitzer}/IRS surveys.

\begin{figure*}[htb!]
    \centering
	\includegraphics[width=0.49\textwidth]{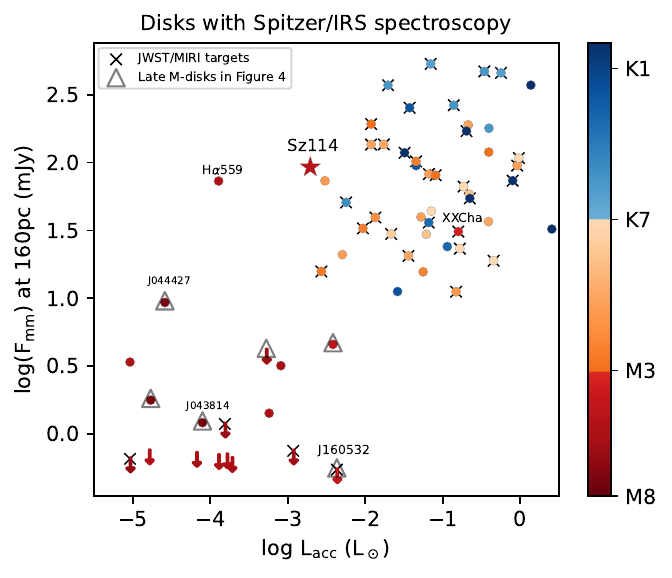}
    \caption{Flux at 0.89\,mm scaled at 160\,pc vs accretion luminosity. Symbols are color-coded by SpTy. The sample of mid-to-late M-star disks includes those in \citet{Pascucci2009ApJ...696..143P} which was only observed at the lowest {\it Spitzer}/IRS resolution. Down arrows indicate millimeter upper limits. Triangles are for the  mid-to-late M-star disks present in Figure~\ref{fig:fluxratio} while crosses indicate JWST/MIRI targets. Sz~114 is indicated with a star symbol and shows a significantly higher millimeter flux than other mid-to-late M-star disks observed with {\it Spitzer}/IRS. } 
    \label{fig:LmmLacc}
\end{figure*}

To place the Sz~114 system into context, we have also collected the accretion luminosity ($L_{\rm acc}$) and the millimeter flux ($F_{\rm mm}$) for the earlier spectral type sample \citep{Banzatti2020ApJ...903..124B} and an expanded late M-star disk sample that now includes systems observed at the lowest {\it Spitzer}/IRS resolution (R$\sim$100) from \cite{Pascucci2009ApJ...696..143P}, see Figure~\ref{fig:LmmLacc}. This latter sample had, on average, a higher C$_2$H$_2$/HCN flux ratio than disks around earlier-type stars but nothing could be said about the strength of the water emission due to the very low spectral resolution. $L_{\rm acc}$ and $F_{\rm mm}$ from this expanded sample are from  \citet{Manara22} which homogenize measurements from various surveys \cite[e.g.,][]{Andrews13, Ansdell16,Pascucci2016ApJ...831..125P, Acala17}.
The reason for focusing on these two parameters is that multiple {\it Spitzer} studies have shown that water emission is positively correlated with accretion luminosity \citep[e.g.,][]{Salyk2011ApJ...731..130S,Banzatti2020ApJ...903..124B} while the water/HCN ratio is anti-correlated with the disk millimeter flux \citep{Najita13}. 
\cite{Banzatti2020ApJ...903..124B} also found the water/HCN ratio, as well as the water luminosity corrected for the $L_{\rm acc}$ trend, is  anti-correlated with dust disk radius as millimeter fluxes are correlated with dust disk sizes \citep[e.g.,][]{Tripathi17,Hendler2020ApJ...895..126H}.

The positive correlation between water and accretion luminosity is likely driven by the additional heating that accretion can provide for the excitation of the lines \citep[e.g.,][]{Najita17,Glassgold2009ApJ...701..142G}. Indeed, the luminosity of several other infrared lines is also positively correlated with $L_{\rm acc}$ (e.g., \citealt{Salyk2011ApJ...731..130S} and Fig.~\ref{fig:NeIILacc}). The  anti-correlation between the water and millimeter flux (and dust disk size) has been interpreted in the context of pebbles' drift: low millimeter fluxes might result from the efficient inward drift of icy pebbles (roughly the mm-sized grains detected by ALMA) which increase the water vapor in the inner disk upon crossing the snowline \citep{Banzatti2020ApJ...903..124B}. This scenario is supported by new MIRI spectra showing excess cold water emission in compact disks, which could originate from enhanced ice sublimation due to a larger flux of inward drifting pebbles \citep{Banzatti2023arXiv230703846B}.

As seen in Figure~\ref{fig:LmmLacc}, Sz~114 has a relatively large accretion luminosity for its spectral type, further supporting the expectation that extra heating boosts infrared lines, including from H$_2$O. However,  Sz~114 and J160532 do not follow the anti-correlation between water luminosity and $F_{\rm mm}$: The infrared spectrum of the millimeter brighter, larger, and structured disk of Sz~114 is significantly more water-rich than that of the millimeter-faint J160532. Note that the spectral type, as well as the stellar mass and accretion luminosity of these two objects are strikingly similar (see Figure~\ref{fig:LmmLacc}). 
One major difference, in addition to the millimeter flux (and potential size and substructures), is the likely age of the two objects: Sz~114 belongs to the $\sim 1-3$\,Myr-old Lupus star-forming region \citep{Galli2020} and has an isochronal age of $\sim 1$\,Myr \citep{Acala17}, while J160532 belongs to the $\sim 5-10$\,Myr-old Upper~Sco association  \citep[e.g.][]{Preibisch08,Luhman20} and has a dynamical age of  2.5$\pm$1.6\,Myr \citep{MiretRoig22}. With the same spectral type but a stellar luminosity that is $\sim 5$ times lower than that of Sz~114, J160532 is clearly older.

How might age play a role in explaining the different mid-IR spectra? Recently, \cite{Mah23}  carried out simulations using a coupled disk evolution and planet formation code that includes pebble drift and evaporation at icelines in radially smooth disks (that is a disk with no gaps or pressure bumps). They showed that, for such disks, the C/O ratio inside the  snowline initially decreases below the stellar value, due to the inward drift and evaporation of water-ice-rich pebbles, before it increases to superstellar values due to the inward diffusion of carbon-rich vapor, mostly from the evaporation of CH$_4$, from the accreting gas surface flow. Importantly, the transition from low to high C/O ratios (water-rich to water-poor) occurs faster around very low-mass stars because of the closer-in icelines \citep[e.g.,][]{Mulders15}, shorter viscous timescales, and faster radial drift \citep[e.g.,][]{Pinilla13}. Specifically, \citealt{Mah23} estimated that the inner disk of a 0.1\,M$_\odot$ star dries out in $\sim$0.5\,Myr if the $\alpha$ viscosity is 10$^{-3}$ and in $\sim 2$\,Myr for $\alpha=10^{-4}$ while that of a 0.7\,M$_\odot$ star dries out in $\sim$2\,Myr and $\sim$10\,Myr, respectively (see their Figure~1). 
This scenario explains the general trend that the inner disks of late M-stars are relatively more water-poor (C-rich) than similarly aged disks around earlier-type stars as well as the difference in the spectra of Sz~114 and  J160532. By the relatively later age of J160532, the icy pebble supply from the outer disk has been likely consumed (as suggested by the low millimeter flux) and the inner disk C/O ratio has been enhanced by the inner diffusion of carbon-rich vapor, which then triggers the formation of hydrocarbons. In contrast, at the early time of Sz~114, the inward drift of icy pebbles and the water vapor released when crossing the snowline, in combination with accretional heating, may boost the water emission. Interestingly, the two tentative detections of water in {\it Spitzer}/IRS spectra of similarly late M-star disks are also from young systems in the Taurus star-forming region \citep[][]{Pascucci2013ApJ...779..178P}. 

\begin{figure*}[htb!]
    \centering
	\includegraphics[width=0.49\textwidth]{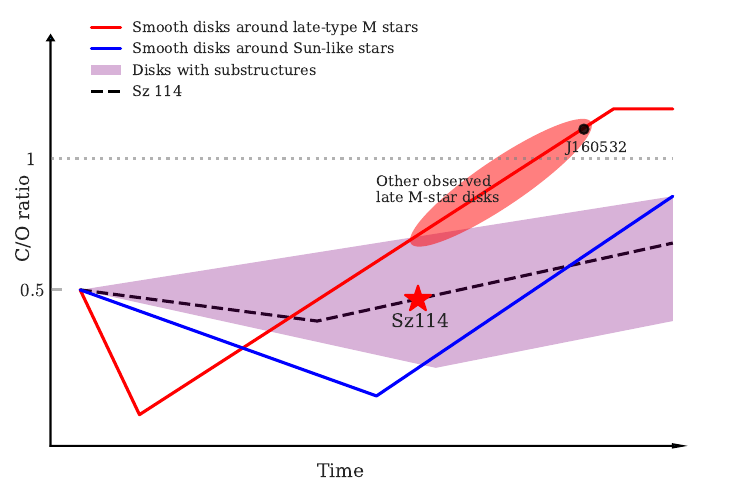}
    \caption{Sketch of the inner disk C/O ratio evolution with time. Substructures, depending on their locations and depths, can slow down the inward drift of icy pebbles. The time at which the C/O ratio starts to increase indicates the depletion time of the inner water-rich phase and depends on several factors including stellar mass, initial disk size, mass, and viscosity, and the presence and depth of substructures. Model predictions for disks with substructures around solar-type stars can be found in \cite{Kalyaan2023ApJ...954...66K}. Models for disks without substructures around stars of different masses are presented in \cite{Mah23}.}
    \label{fig:sketch}
\end{figure*}

While these ideas may account for the relative properties of Sz~114 and J160532, it is also interesting to consider why Sz~114 is so different from the majority of the mid-to-late M-star disks of similar age observed with {\it Spitzer}/IRS.  
One possibility is that Sz114 formed with unusual initial conditions, i.e. a large and massive disk at birth. Sz~114 has an exceptionally large millimeter flux and extended dust disk among mid-to-late M-stars. 
Sz~114 may also be unusual among mid-to-late M-star disks in hosting substructures (gaps).  The shallow gap detected by DSHARP at $\sim 39$\,au, along with potential other  substructures between $\approx 7-12$\,au, might maintain some icy pebbles further out, thus maintaining a large disk, and, at the same time, prolong water enrichment in the inner disk (see e.g. Fig.~4 in \citealt{Kalyaan2023ApJ...954...66K} for this effect in disks around solar-type stars). 
In contrast, most mid-to-late M-star disks are characterized by low millimeter fluxes 
(e.g., \citealt{Pascucci2016ApJ...831..125P} and Figure~\ref{fig:LmmLacc} for those targeted with {\it Spitzer}/IRS). These faint disks are low in the total amount of icy pebbles: they might have started with an overall smaller and less massive disk than Sz~114 and have not developed any substructure to slow down the inner drift of icy pebbles. 
As a result, the typical inner disk of late M-stars would follow the evolution described in \cite{Mah23} and ``dry out'' fast as discussed above. A sketch that places the temporal evolution of the C/O ratio in context, for both Sz 114 and other late M-star disks, is provided in Figure~\ref{fig:sketch}. 


These ideas can be tested with MIRI-MRS spectroscopy of larger samples of disks surrounding mid-to-late M stars. JWST GTO and GO Cycle~1 and 2 programs will also cover the low millimeter flux disks around late M-stars (see crosses in Figure~\ref{fig:LmmLacc}) for which, similar to J160532, we expect water-poor and C-rich spectra even at young ages. It would be valuable to carry out MIRI-MRS observations of the  late M-star  disk in Taurus with tentative water detection and moderately high F$_{\rm mm}$ (J044427), as well as of ESO~H$\alpha$~559, which has a  millimeter flux comparable to Sz~114. 
If our interpretation of the data at hand is correct, we expect these spectra to be water-rich as the spectrum of Sz~114.  Fully testing the proposed scenario would require 
 sensitive MIRI-MRS spectra of disks spanning a large range in spectral type, millimeter fluxes,  accretion luminosities, and ages across several star-forming regions. 


\section{Summary and Outlook} \label{sec:summ}
We have acquired and analyzed the JWST/MIRI-MRS spectrum of Sz~114, an accreting M5 star in the $\sim 1-3$\,Myr-old Lupus star-forming region surrounded by a Class~II disk. This system has a high millimeter flux and was selected for the ALMA Disk Substructures at High Angular Resolution Project (DSHARP), which discovered a large dust disk with a size of $\sim 58$\,au and a shallow gap at $\sim 39$\,au. Additional unresolved  substructures may be also present between $\approx 7-12$\,au \citep{Jennings22}. This disk was not previously observed with {\it Spitzer}/IRS at $R \sim 700$, hence there are no gas line detections previously reported. We model the MRS spectrum of Sz~114 with a  slab of gas in LTE with the main goal of removing line blending and calculating fluxes for the detected gas lines.
We also collect {\it Spitzer}/IRS fluxes from  disks around stars of similar and earlier spectral type and discuss the spectrum of Sz~114 in the context of this larger sample. Our main results can be summarized as follows:
\begin{itemize}
\item The MIRI-MRS spectrum of Sz~114 shows a forest of water emission lines, from the hot water bending rovibration band at $\sim 6.6$\,\micron{} through  the cooler pure rotational lines at $\sim 25$\,\micron .
In addition to water, we also detect CO, CO$_2$, HCN, and C$_2$H$_2$, with  CO$_2$ having the strongest ro-vibrational band. Two H$_2$ lines and the \neii{} at 12.81\,\micron\ are also robustly detected while $^{13}$CO$_2$ is only marginally detected. 
\item The water spectrum of Sz~114 between $\sim 10-19$\,\micron\ requires at least two LTE slabs of gas: hot close-in ($\sim 800$\,K) water emission and an additional cooler ($\sim 500$\,K)  component whose emission comes from further out in the disk.
\item The 17.22\,\micron{} water flux of Sz~114 exceeds that of mid-to-late M-star disks, including that of the carbon-rich disk of J160532 observed with JWST, by an order of magnitude, and falls within the range observed in  disks around earlier-type stars by {\it Spitzer}. Similarly, the C$_2$H$_2$/H$_2$O and HCN/H$_2$O flux ratios of Sz~114 align with those of disks around earlier-type stars, while exhibiting a slightly higher CO$_2$/H$_2$O ratio.
\end{itemize}

As for the earlier-type stars, most of the molecular emission from Sz~114 is expected to arise in the disk atmosphere inside the snowline.  
While accretion luminosity can boost molecular emission, it alone cannot account for the rich water spectrum observed in Sz~114: J160532, for example, exhibits a similar $L_{\rm acc}$ but a water-poor and C-rich spectrum. 
Several other factors might contribute to the exceptionally rich water spectrum of this late M-star disk. We have highlighted age because the water-rich inner disk phase is much shorter in smooth disks around lower-mass stars: This is because 
icy pebbles, which are O-rich, are depleted faster and the inner disk C/O ratio is enhanced earlier through the diffusion of C-rich vapor \citep{Mah23}. Age could explain the difference between the MIRI spectra of Sz~114 and the relatively older J160532. In addition, we propose that unusual initial conditions, a large massive disk at birth, combined with the development of substructures, might be necessary to understand Sz~114 in the context of other mid-to-late M-star disks of similar age observed with {\it Spitzer}/IRS. The existence of the $\sim 39$\,au shallow gap in the disk of Sz~114   might help to retain a large dust size but it is not a strong enough barrier to pebbles drifting inward and could just help prolong the water-rich phase of the inner disk \citep{Kalyaan2023ApJ...954...66K}. 
The other inferred substructures between $\approx 7-12$\,au might further delay the delivery of water vapor inside the midplane snowline which is expected to be only at $\sim 0.1-0.5$\,au from the star. This delay would preserve a lower C/O ratio for a more extended period compared to other mid-to-late M-star disks, the majority of which are millimeter faint and likely lack significant dust traps. 
The lower frequency of dust traps, in combination with a typically low disk mass at birth, could have important implications for the type of planets forming  in disks around late M-stars. For instance, in the context of planet formation via pebble accretion, \cite{Mulders2021ApJ...920L...1M} predict that the frequency of close-in super-Earths should decrease for M$_* \le 0.5$M$_\odot$, meaning that systems like TRAPPIST-1 would be rare.

Our findings demonstrate that the mid-infrared spectra of mid-to-late M-star disks exhibit greater diversity than previously inferred from the limited subset observed with {\it Spitzer}. This underscores the importance of expanding the sample size to comprehensively explore and better understand the potential range of variations.

\begin{acknowledgments}
This work is based on observations made with the NASA/ ESA/CSA James Webb Space Telescope. The data were obtained from the Mikulski Archive for Space Telescopes at the Space Telescope Science Institute, which is operated by the Association of Universities for Research in Astronomy, Inc., under NASA contract NAS 5-03127 for JWST. The observations are associated with the JWST GO Cycle~1 program 1584.
C.X. and I.P. acknowledge partial support by NASA under Agreement No. 80NSSC21K0593 for the program ``Alien Earths”.
Support for F.L. was provided by NASA through the NASA Hubble Fellowship grant \#HST-HF2-51512.001-A awarded by the Space Telescope Science Institute, which is operated by the Association of Universities for Research in Astronomy, Incorporated, under NASA contract NAS5-26555. A portion of this research was carried out at the Jet Propulsion Laboratory, California Institute of Technology, under a contract with the National Aeronautics and Space Administration (80NM0018D0004). Some/all of the data presented in this paper were obtained from the Mikulski Archive for Space Telescopes (MAST) at the Space Telescope Science Institute. The specific observations analyzed can be accessed via \dataset[DOI]{https://doi.org/10.17909/adqk-w250}.

\end{acknowledgments}

%

\vspace{5mm}
\facilities{JWST(MIRI MRS), ALMA, {\it Spitzer}(IRS)}


\software{{\tt astropy} \citep{2013A&A...558A..33A,2018AJ....156..123A}, {\tt scipy} \citep{Virtanen20} }



\appendix
\section{Fitting process and error bars}\label{sec:Appendix:errorbars}
We apply a custom-written  random walk process to fit the MIRI spectrum of Sz~114. The likelihood function is defined as follows \citep[e.g.,][]{Hogg10}{}{}: 
\begin{equation}
    \ln(\mathcal{L})=-\frac{1}{2} \sum_{\lambda} \left[\left(\frac{F_{M}(\lambda, \theta)-F_D(\lambda)}{s_D}\right)^2+\ln(2\pi s_D^2)\right] 
\end{equation}
The $F_M(\lambda,\theta)$ is the flux of our model at  $\lambda$ with the parameter set $\theta(N,R_{\rm em},T)$ and $F_D(\lambda)$ is the continuum-subtracted JWST MIRI data. 
$s_D^2= \sigma_D^2+ f^2(F_M(\lambda,\theta))^2$ indicates the variance of our data \citep{Hogg10}, where $\sigma_D$ is the uncertainty of the data and $f\times F_M(\lambda, \theta)$ is a factor accounting for a possible underestimation of the uncertainty. We simplify this process by ignoring this possible underestimation, i.e. we assume $f=0$, and the likelihood is therefore simplified to $\ln(\mathcal{L})=K-\frac{1}{2}\chi^2$ where $K=-\frac{1}{2} \sum_{\lambda} \ln(2\pi \sigma_D^2)$ is a constant and $\chi^2=\sum_{\lambda} \left(\frac{F_{M}(\lambda, \theta)-F_D(\lambda)}{\sigma_D} \right)^2$. We also apply a flat prior inside the boundaries mentioned in Section~\ref{sec:water-rich} throughout our fitting. 

\begin{figure*}[htb!]
    \centering
	\includegraphics[width=0.49\textwidth]{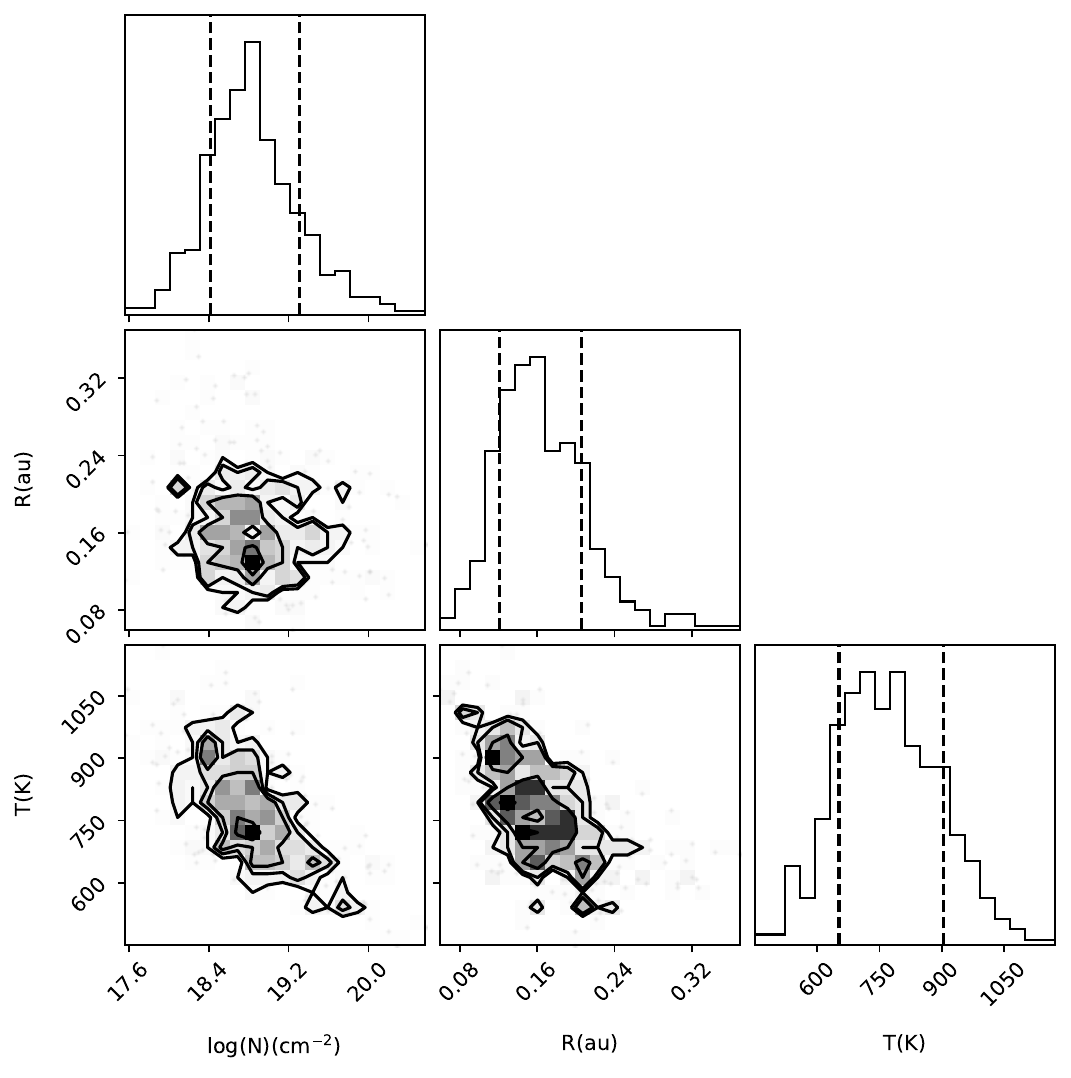}
    \caption{Corner plot of our hot water models as an example of our fitting. The vertical dashed lines indicate the $1\sigma$ of each parameters as the middle 68$\%$ of the accepted models. We can see the degeneracy between the three parameters from the shape of the distribution. 
    }
    \label{fig:cornerplot}
\end{figure*}

Our random walk process works as follows: For each initial parameter set $\theta(N,R_{\rm em},T)$ we run a model with CLIcK and compare our model with the data to get a likelihood $\mathcal{L(\theta)}$. Then we `walk' the parameter set $\theta$ to 8 different directions as $\theta_i(N\times \Delta N_i,R_{\rm em}+\Delta R_{\rm em, i},T+\Delta T_i)$ and calculate the likelihood for each of them $\mathcal{L(\rm \theta_i)}$. The model with the largest likelihood among the 8 models is chosen to be an `accepted model' and become the next initial parameter set. We repeat the process until the likelihood function settles to a certain value. 
The best-fit model is the model with the highest likelihood. This approach is similar to previous grid fittings which minimize $\chi^2$ \citep[e.g.,][]{Tabone2023NatAs.tmp...96T,Grant2023ApJ...947L...6G}{}{}. The error bars are determined through typical Markov Chain Monte Carlo (MCMC) methods, where $1\sigma$ uncertainties are chosen as the 16th and 84th percentile values from the distribution of parameter space spanned by our walkers (see Figure~\ref{fig:cornerplot}).
Due to the overlapping emission, C$_2$H$_2$ and HCN are fitted together. The column density (N) and temperature (T) of the two molecules are highly degenerate with each other, thus the error bars of the fitted parameters are very large and not shown here
(e.g., table~\ref{tab:model}). We note that our major goal is to separate the contribution from different molecules to calculate line fluxes, and use these strategies mentioned above to speed up our modelling. See Figure~\ref{fig:compositemodel}, \ref{fig:expandedview} for the comparison between our fitted models and the data. 

\begin{deluxetable}{lccccc}[thb!]
\tablecaption{Summary of model fitting\label{tab:model}}
\tablewidth{0pt}
\tablehead{
\multicolumn{2}{c}{Species}&  \colhead{$\lambda-$range}  & \colhead{$T$} & \colhead{$\log_{10}(N)$} &\colhead{$R_{\rm em}$}   \\
\colhead{} &\colhead{}& \colhead{($\mu$m)} & \colhead{(K)} & \colhead{(cm$^{-2}$)}  & \colhead{(au)} 
}
\startdata
\multirow{2}{2em}{H$_2$O} & hot & 10-12& 750$^{+150}_{-130}$ & 18.8$\pm$0.4 & 0.17\\
& cool & 16.3-19& 450$^{+90}_{-60}$ & 18.2$^{+0.8}_{-0.3}$ & 0.5 \\
\multicolumn{2}{c}{CO$_2$}   & 14.6-15.4 & 500$^{+200}_{-150}$ & 17.6$^{+1.2}_{-1.1}$ & 0.18\\
\multicolumn{2}{c}{C$_2$H$_2$} & 13.5-14.3 & 1400 &15.5  &0.42\\
\multicolumn{2}{c}{HCN}  & 13.5-14.3 & 870 & 15.9  &0.39 \\
\enddata
\tablecomments{We identify two sets of well-fit parameters for C$_2$H$_2$ and HCN given the intrinsic degeneracies between column density, temperature, and emitting area in LTE slab models of warm gas in protoplanetary disks \citep[e.g.,][]{Salyk2011ApJ...731..130S,Najita2018ApJ...862..122N}{}{}. The values reported here correspond to the maximum likelihood in our fitting routine, and the other parameter set is $T \sim$ 400~K and log$_{10}(N) \sim 19$ cm$^{-2}$ for both C$_2$H$_2$ and HCN. 
}
\end{deluxetable}

\section{\neii\ and H$_2$ emission}\label{app:OtherLines}
As mentioned in Sect.~\ref{sec:water-rich}, we have detected two molecular hydrogen transitions, both with peak centroids consistent with the stellar radial velocity: the S(3) at $\sim 9.67$\,\micron\ and the S(1) at $\sim 17.035$\,\micron. The critical densities of these two lines are low \citep{Mandy93} compared with the typical density of protoplanetary disks in the H$_2$ emitting region  \citep[e.g.,][]{Woitke2009A&A...501..383W}. Therefore, the population levels are likely to be in LTE. With the further assumption that the emission is optically thin, we use eq.~1  in \cite{Lahuis07} and derive a temperature of $\sim 425$\,K from their line ratio,  which is consistent with temperatures near the disk surface \citep[e.g.,][]{Kamp04,Woitke18}.

\begin{figure*}[htb!]
    \centering
	\includegraphics[width=0.49\textwidth]{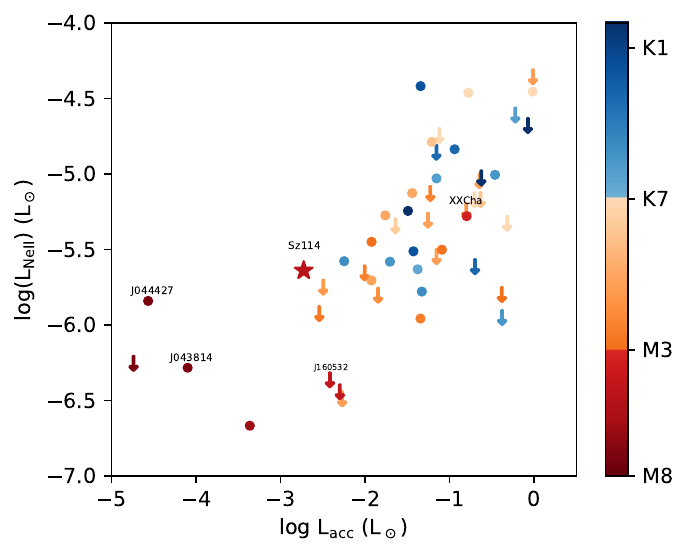}
    \caption{\neii{} vs. accretion luminosities for the sample of late M-type  \citep{Pascucci2013ApJ...779..178P} and disks around earlier-type stars \citep{Banzatti2020ApJ...903..124B}. For the latter sample \neii{} fluxes are collected from \cite{Guedel2010A&A...519A.113G} and \cite{Rigliaco2015ApJ...801...31R}.
    }
    \label{fig:NeIILacc}
\end{figure*}

Sz~114 has also a fairly strong \neii\ 12.81\,\micron\ detection for its spectral type.  In Figure~\ref{fig:NeIILacc} we compare its line luminosity to that of other late M-star disks   \citep{Pascucci2013ApJ...779..178P} and disks around earlier-type stars \citep{Banzatti2020ApJ...903..124B}
as a function of accretion luminosity ($L_{\rm acc}$).  \neii{} fluxes for the disks around earlier-type stars are collected from \cite{Guedel2010A&A...519A.113G} and \cite{Rigliaco2015ApJ...801...31R} and luminosities are computed using the most updated Gaia DR3 distances \citep{Gaia3}. 
Sz~114 is not only bright in the \neii\ line for its spectral type but also brighter than the average for sources with the same $L_{\rm acc}$. This hints at extra \neii{} emission from a jet \citep[e.g.,][]{Guedel2010A&A...519A.113G} which is supported by the 70\,km s$^{-1}$  blueshift in the peak centroid with respect to stellar radial velocity.  
A high-velocity blueshifted component is also detected in the \oi\ 6300\,\AA\ profile obtained with the X-Shooter/VLT \citep{Nisini2018A&A...609A..87N}. A visual inspection of the MIRI-MRS datacube around the \neii{} line does not show any extension possibly due to the almost face-on (21$^\circ$, \citealt{Huang18}) disk inclination.  More detailed analysis of the MIRI-MRS datacubes and/or higher S/N exposures might reveal a spatially resolved \neii{} jet and enable measuring the proportion of jet versus disk emission.

\bibliography{ref}
\bibliographystyle{aasjournal}



\end{document}